\documentclass{iau}
\usepackage{graphicx} 

\title[IAUS291.~~Pulsar Timing Arrays] %% short title %%
{Pulsar Timing Arrays: Status and Techniques} %% full title %%

\author[G. Hobbs]  %% short author list %%
{George Hobbs$^1$
}

\affiliation{$^1$CSIRO Astronomy and Space Science, Australia Telescope National Facility, PO~Box~76, Epping
NSW~1710, Australia \\ email: {\tt george.hobbs@csiro.au} \\[\affilskip]}

\pubyear{2012}
\volume{291}  %% insert here IAU Symposium No.
\jname{\mbox{Neutron Stars and Pulsars: Challenges and Opportunities after 80 years}}
\editors{J.~van Leeuwen, ed.} 
\begin{document}
\maketitle

%% -- Abstract ----------------------------------
\begin{abstract}
Three pulsar timing arrays are now producing high quality data sets.  As reviewed in this paper, these data sets are been processed to 1) develop a pulsar-based time standard, 2) search for errors in the solar system planetary ephemeris and 3) detect gravitational waves.   It is expected that the data sets will significantly improve in the near future by combining existing observations and by using new telescopes.
\end{abstract}

% add below any authors, subjects and objects for indexing 
%   add more lines if necessary
%   but leave all lines commented out
%\index[author]{LastName1, Initials|textbf}
%\index[author]{LastName2, Initials|textbf}
%\index[subject]{Keyword1}
%\index[subject]{Keyword2}
%\index[object]{Object1}
%\index[object]{Object2}

\firstsection % if your document starts with a section,
\section{Introduction}
Pulsar timing data sets are now of sufficient length and precision to start to realise many of the goals of ``pulsar timing arrays'' (PTAs).   The first major PTA was initiated in 2004 using the Parkes radio telescope (Manchester et al., 2012) and is known as the Parkes Pulsar Timing Array (PPTA) project.   The European PTA (EPTA) makes use of the telescopes at Jodrell Bank, Westerbork, Effelsberg, Nan\c cay and Sardinia (e.g., Ferdman et al. 2010).  The North American PTA (NANOGrav; Jenet et al. 2009) obtains observations using the Arecibo and Green Bank radio telescopes.  Together these three PTAs form the International Pulsar Timing Array (IPTA; Hobbs et al. 2010a) which provides high quality timing observations of approximately 40 of the most stable millisecond pulsars known.  

The first stage of PTA-related data analyses is to determine the pulse times-of-arrival (ToAs) for each observation of each pulsar.   These ToAs are converted from the observatory time standard to a realisation of terrestrial time (TT).  Barycentric arrival times are calculated using knowledge of the relative position of the Earth with respect to the solar system barycentre using a planetary ephemeris and by converting from TT to coordinate barycentric time (TCB).  These barycentric arrival times are compared with predictions of the arrival times using a model for the pulsar rotational and orbital parameters.  The differences between the actual measurements and the predictions are known as the ``pulsar timing residuals''.  This technique, known as ``pulsar timing", is widely used in pulsar astronomy and is described in detail by Edwards, Hobbs \& Manchester (2006).

Various phenomena such as gravitational waves, unexplained timing irregularities, glitch events, errors in terrestrial time standards or in the solar-system ephemeris will induce timing residuals. The main aim of PTAs is to distinguish between these various phenomena by searching for correlations between the timing residuals of multiple pulsars.  For instance, pulsar timing irregularities, glitch events or interstellar medium variations will lead to timing residuals that are uncorrelated between different pulsars.   In contrast an error in the terrestrial time standard will lead to timing residuals that are identical for different pulsars (assuming that all pulsars have been observed over the same time span).  Errors in the solar-system ephemeris will affect different pulsars depending upon their ecliptic coordinates.   The phase and amplitude of timing residuals induced by a gravitational wave will depend upon the pulsar-Earth-source angle (e.g., Detweiler 1979).  The expected correlation between different pulsars for timing residuals induced by an isotropic, stochastic gravitational wave background has been calculated by Hellings \& Downs (1983).

\section{Developing a pulsar time standard}

Millisecond pulsar rotation is incredibly stable.  This leads to the possibility of developing a time scale based on the pulsar rotation analogous to the free atomic scale, \'Echelle  Atomique Libre (EAL). The Ensemble Pulsar Scale (EPS) can be used to detect fluctuations in atomic timescales and therefore can lead to a new realisation of Terrestrial Time, TT.

\begin{figure}
\begin{center}
\includegraphics[angle=-90,width=10cm]{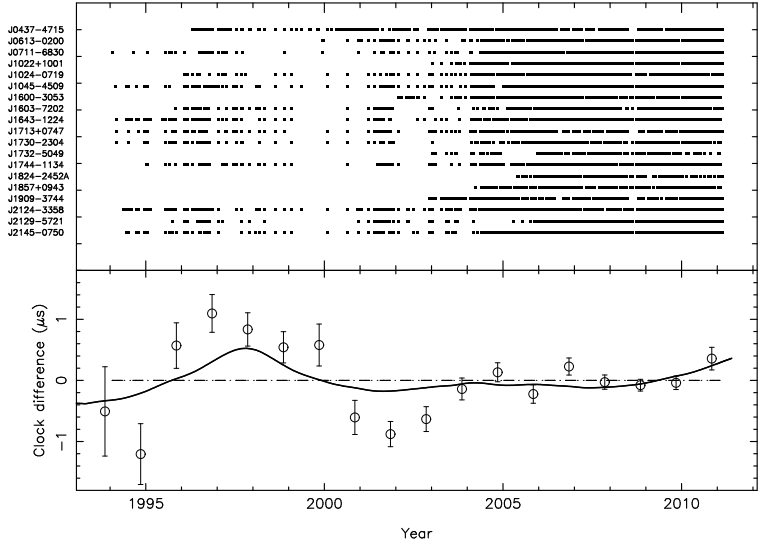}
\end{center}
\caption{This figure is reproduced from Hobbs et al., (2012). The top panel shows the sampling for each of the pulsars in our sample.  The lower panel shows the difference between the pulsar timescale and TT(TAI) as points with error bars.  The solid line indicates the difference between TT(TAI) and TT(BIPM11) after a quadratic polynomial has been fitted and removed.  Full details are available in Hobbs et al. (2012).}
\label{fg:clock}
\end{figure}

Earlier attempts to develop a pulsar timescale have been made by Guinot \& Petit (1991)\nocite{gp91}, Petit \& Tavella (1996)\nocite{pt96},
Rodin (2008)\nocite{rod08} and Rodin \& Chen (2011)\nocite{rc11}.   Recently we have developed a method to produce a new time scale based on observations of 19 pulsars obtained for the PPTA project (Hobbs et al. 2012).  The new algorithm has been implemented as part of the \textsc{tempo2} software package (Hobbs, Edwards \& Manchester 2006).  This algorithm accounts for various features of the observations such as: 1) irregular sampling, 2) different data spans for different pulsars and 3) different fitting parameters for different pulsars.  Our result is reproduced in Figure~\ref{fg:clock}.  We successfully follow features known to affect the frequency of the International Atomic Timescale (TAI)  and we find marginally significant differences between our pulsar time scale, TT(PPTA11), and TT(BIPM11). 

This work is being continued by combining the Parkes observations with data from other observatories.  The new analysis will confirm or deny the tentative discrepancies between TT(PPTA11) and TT(BIPM11) whilst significantly improving the stability and precision of the pulsar scale.  In the longer term it is expected that a future pulsar time scale will be combined with the best atomic timescale to give the world's most stable time scale that will be valid effectively forever. 

\section{Improving the solar system ephemeris}

\begin{figure}
\begin{center}
\includegraphics[angle=-90,width=10cm]{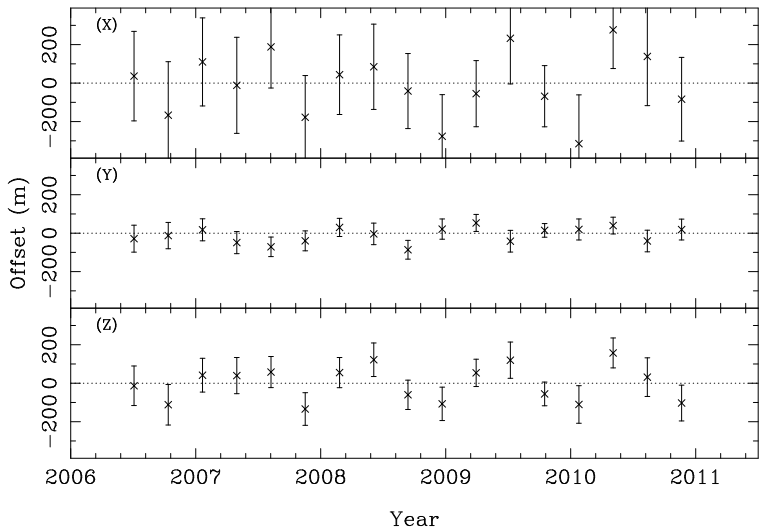}
\end{center}
\caption{Offset of the Earth-Solar System barycentre vector as a function of time from the JPL DE421 planetary ephemeris. The three panels show the offset in the three spatial coordinates.}
\label{fg:ephem}
\end{figure}

The pulsar timing method relies on the  determination of pulse times of arrival as measured in the solar-system barycentre.  The procedure requires knowledge of the position of the Earth with respect to the solar-system barycentre. This is obtained using a published solar system ephemeris. Errors in the ephemeris will lead to timing residuals.   For instance, an error in the mass of Jovian system assumed when forming the ephemeris will lead to residuals proportional to both the pulsar-barycentre-Jupiter angle and the size of the mass error.  As Jupiter orbits the barycentre, the angle will change and hence sinusoidal pulsar timing residuals will be induced with a period equal to that of Jupiter's orbit.

Champion et al. (2010) searched for such sinusoidal timing residuals using PTA observations of four pulsars obtained using the Arecibo, Parkes and Effelsberg radio telescopes.   In most cases published masses obtained from space-craft data were more precise than the pulsar results.  However, for the Jovian system, the Champion et al. (2010) measurement of $9.547921(2) \times 10^{-4}$\,M$_\odot$ is  more accurate than the mass determined from the \emph{Pioneer} and \emph{Voyager} space-craft.

The Champion et al. (2010) technique can only be applied to known solar-system objects. However, it is also possible to determine offsets from the predictions of a specific ephemeris in the Earth's position with respect to the barycentre.  Significant offsets in any of the three spatial coordinates can subsequently be analysed in order to identify the orbital parameters of any previously unknown object.  In Figure~\ref{fg:ephem} we show the results of an initial analysis using the PPTA observations.  We plot the offset in the Earth-barycentre vector as a function of time compared with the value predicted using the JPL DE421 Solar System ephemeris.  We identify no significant offsets suggesting that the ephemeris is adequate for our current purposes over our five year data span.

The sensitivity of a PTA to errors in the solar system ephemeris depends upon the timing precision achieved and the data span.  It is expected that significant improvements will occur when the data span becomes longer than 29 years, the orbital period of Saturn.  Combining the observations from the existing timing arrays will also significantly improve our sensitivity.

\section{Searching for gravitational waves}

\begin{figure}
\begin{center}
\includegraphics[angle=-90,width=10cm]{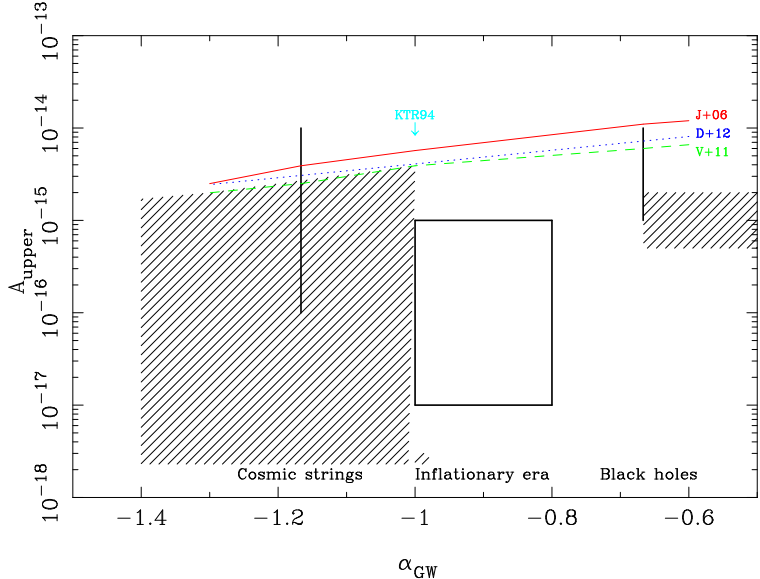}
\end{center}
\caption{Current upper bounds on the gravitational wave background and past and current predictions. The different symbols are shown in the text.}
\label{fg:limit}
\end{figure}

Sazhin (1978) and Detweiler (1979) showed that gravitational waves (GWs) passing through the solar system will induce timing residuals that are potentially detectable using PTAs.  Pulsar data sets are sensitive to GWs with periods longer than the typical data sampling and shorter than the total time span of the observations.  Hence, pulsar experiments are sensitive to ultra-low frequency ($\sim 10^{-9}$--$10^{-8}$\,Hz) GWs and are complementary to ground-based and space-based GW detectors such as LIGO and eLISA.

Sources of a background of GWs include cosmic strings (see Sanidas et al. 2012, Regimbau et al. 2012 and references therein), the inflationary era (e.g., Zhao 2011) and coalescing supermassive binary black holes (e.g., Sesana, Vecchio \& Colacino 2008).  The induced timing residuals induced by a GW background is often described as having a red power spectrum:
\begin{equation}
P(f) = \frac{A^2}{12\pi^2}\left(\frac{f}{f_{\rm 1yr}}\right)^{2\alpha_{GW}-3}.
\end{equation} 
The almost horizontal lines in Figure~\ref{fg:limit} show recent upper bounds that have been placed on $A(\alpha)$ by Jenet et al., (2006), Van Haasteren et al. (2011) and Demorest et al. (2012).  For comparison an earlier upper bound by Kaspi, Taylor \& Ryba (1994) is plotted as a downward pointing arrow.  

Since the start of the PPTA project our expectations for the detectable GW signal have changed. In Figure~\ref{fg:limit} we show the past and current predictions for the signal strength and spectral exponent. Estimates that were available at the time that our first upper bound was published (Jenet et al. 2006) for cosmic strings, the inflationary era and black holes are shown as the left-most vertical solid line, the hollow rectangle and the right-most vertical solid line respectively. Since 2006, new research has (1) broadened the possible range of $\alpha_{\rm GW}$ and indicated that very few constraints are available on the lower amplitude of the cosmic-string background signal (Sanidas, Battye \& Stappers 2012; shown as the left-most hashed region in Figure~\ref{fg:limit}), (2) confirmed that the most-likely signal from the inflationary era has a very low amplitude (the small hashed region near the bottom-centre of the figure should be considered as an upper-bound on the amplitude of this signal) and 3) led to the possibility that the background caused by coalescing black holes is flatter than originally predicted (right-most hashed region).

Earlier work has assumed that the detected GW signal will be an isotropic, stochastic background.  More recent studies have suggested that  the exact nature of the detected signal could be a background, but may be an individual non-evolving source, a chirping system, a memory event, or a burst event.  Various algorithms have therefore been developed to search for these various signals (e.g., Yardley et al. 2010; Finn \& Lommen 2010; Cordes \& Jenet 2012).

\section{Improving the data sets}

The sensitivity of a PTA data set to a given science goal depends upon the timing precision achieved, the noise present in the data, the number of pulsars observed and the data span.  For most millisecond pulsars over decadal time scales the dominant noise source is caused by turbulence in the interstellar medium leading to variations in the pulsar's dispersion measure (e.g., You et al. 2007).  This effect can only be removed by observing the pulsar at widely separated observing frequencies.  To help address this problem many PTAs are now developing wide-band receiver systems that provide a large frequency coverage for each observation.  

For many pulsars, the dominant uncorrectable noise is caused by receiver noise and pulse jitter.  Receiver noise can only be reduced using more sensitive telescopes or longer observations.  In the future it is expected that telescopes such as the Five-hundred metre spherical telescope (FAST) in China, or the Square Kilometre Array (SKA) will provide a huge increase in sensitivity.  However, individual pulse-shape variations or jitter may provide a limit to the precision with which pulse arrival times can be measured and therefore require a modified strategy in the use of these new telescopes (Oslowski et al. 2011, Liu et al. 2012).

Over long time scales pulsars are known to exhibit irregularities in their spin-down rate (e.g. Verbiest et al. 2009 and Hobbs et al. 2010b).  It is currently thought that this noise is uncorrectable and will limit the stability of pulsars over long time scales.  However, recent work (Lyne et al. 2011) has shown that it may be possible to identify a deterministic component  to these irregularities which opens up the possibility of at least partially correcting for their effects.  In any case, the discovery of new, stable pulsars is necessary to improve the sensitivity of the PTA projects.  Numerous surveys are ongoing (e.g., Cordes et al. 2006, Keith et al. 2010, Boyles et al. 2012) and are leading to the discovery of a large number of new, millisecond pulsars.  Including these new pulsars in existing PTAs and the likelihood of a large number of new, sensitive radio telescopes in the relatively near future suggest that the future is bright for PTA research.

\section{Acknowledgements}

We acknowledge the dedication and skills of the engineers and support staff at the various observatories without whom the various Pulsar Timing Array projects could not exist.  GH thanks A. Sesana, S. Sanidas and X. Siemens for interesting discussions on the parameters of the expected gravitational wave background and R. Manchester, W. Coles and J. Verbiest for comments on the manuscript.

% CUP work flow only accepts EPS -- not PDF, JPG, etc.
% \begin{figure}[b]
% \begin{center}
%  \includegraphics[width=3.4in]{YourFig.eps} 
%  \caption{Path of pre-solar grains from their stellar sources to the
%    laboratory.} 
%    \label{fig1}
% \end{center}
% \end{figure}


\begin{thebibliography}{}

\bibitem[Boyles \etal\ (2012)]{blr+12}
{Boyles, J., et al.} 2012, \textit{arXiV}, 1209, 4293

\bibitem[Champion \etal\ (2010)]{chm+10}
{Champion, D., et al.} 2010, \textit{ApJ}, 720, 201

\bibitem[Cordes \etal\ (2006)]{cfl+06}
{Cordes, J. M. et al.} 2006, \textit{ApJ}, 637, 446

\bibitem[Cordes \etal\ (2012)]{cj12}
{Corde, J. M. \& Jenet, F. A.} 2012, \textit{ApJ}, 752, 54

\bibitem[Demorest \etal (2012)]{dem12}
{Demorest, P. et al.} 2012, \textit{arXiv}, 1201, 6641

\bibitem[Detweiler (1979)]{det79}
{Detweiler, S.} 1979, \textit{ApJ}, 234, 1100

\bibitem[Edwards \etal\ (2006)]{ehm06}
{Edwards, R., Hobbs, G., \& Manchester, R.} 2006, \textit{MNRAS}, 372,  1549

\bibitem[Ferdman \etal\ (2010)]{fvb+10}
{Ferdman, R. D., et al.} 2010, \textit{CQGra}, 27, 4014

\bibitem[Finn \etal\ (2010)]{fl10}
{Finn, L. S. \& Lommen, A. N.} 2010, \textit{ApJ}, 718, 1400

\bibitem[Guinot \etal\ (1991)]{gp91}
{Guinot, B., \& Petit, G.} 1991, \textit{A\&A}, 248, 292

\bibitem[Hellings \etal\ (1983)]{hd83}
{Hellings, R. \& Downs, G.} 1983, \textit{ApJ}, 256, 39

\bibitem[Hobbs \etal\ (2006)]{hem06}
{Hobbs, G., Edwards, R. \& Manchester, R.} 2006, \textit{MNRAS}, 369, 655

\bibitem[Hobbs \etal\ (2010)]{haa+10}
{Hobbs, G., et al.} 2010a, \textit{CQGra}, 27, 4013

\bibitem[Hobbs \etal\ (2010)]{haa+10}
{Hobbs, G., Lyne, A. \& Kramer, M.} 2010b, \textit{MNRAS}, 402, 1027

\bibitem[Hobbs \etal\ (2012)]{hcm+12}
{Hobbs, G. et al.} 2012, accepted by MNRAS (arXiv.1208.3560)

\bibitem[Jenet \etal\ (2006)]{jhv+06}
{Jenet, F. et al.} 2006. \textit{ApJ}, 653, 1571

\bibitem[Jenet \etal\ (2009)]{jfl+09}
{Jenet, F. et al.} 2009. \textit{arXiv}, 0909.1058

\bibitem[Kaspi \etal\ (1994)]{ktr94}
{Kaspi, V. et al.} 1994. \textit{ApJ}, 428, 713

\bibitem[Keith \etal\ (2010)]{kjv+10}
{Keith, M. J. et al.} 2010. \textit{MNRAS}, 409, 619

\bibitem[Liu \etal\ (2012)]{lkl12}
{Liu, K. et al.} 2012, \textit{MNRAS}, 420, 361

\bibitem[Lyne \etal (2010)]{lhk+10}
{Lyne, A. et al.} 2010, \textit{Sci}, 329, 408

\bibitem[Manchester \etal (2012)]{m+12}
{Manchester, R. N. et al.} 2012, \textit{submitted to MNRAS}

\bibitem[Oslowski, S. \etal\ (2011)]{ovh+11}
{Oslowski, S. et al.} 2011. \textit{MNRAS}, 418, 1258

\bibitem[Petit \etal\ (1996)]{pt96}
{Petit, G. \& Tavella, P.} 1996, \textit{A\&A}, 308, 290

\bibitem[Regimbau \etal\ (2012)]{rgsm12}
{Regimbau, T., Giampanis, S. Siemens, X \& Mandic, V.} 2012, \textit{PhRvD}, 85, 6001

\bibitem[Rodin (2008)]{rod08}
{Rodin, A.} 2008, \textit{MNRAS}, 387, 1583

\bibitem[Rodin \etal \(2011)]{rod11}
{Rodin, A. \& Chen, D.} 2011, \textit{Astronomy Reports}, 55, 622

\bibitem[Sanidas \etal\ (2012)]{sbs12}
{Sanidas, S., Battye, R \& Stappers, W.} 2012,
\textit{PhRvD}, 85, 12003

\bibitem[Sesana \etal\ ()]{ses}
{Sesana, A., Vecchio, A. \& Colacino, C. N.} 2008, \textit{MNRAS}, 390, 192

\bibitem[Sazhin (1979)]{saz79}
{Sazhin, M. V.}, 1978, \textit{Sov. Astron.}, 22, 36

\bibitem[van Haasteren \etal\ (2011)]{vlj+11}
{van Haasteren, R., et al.} 2011, \textit{MNRAS}, 414, 3117

\bibitem[Verbiest \etal\ (2009)]{vbc+09}
{Verbiest, J. P. W., et al.} 2009, \textit{MNRAS}, 400, 951

\bibitem[Yardley \etal\ (2010)]{yhj+10}
{Yardley, D., et al.} 2010, \textsc{MNRAS}, 407, 669

\bibitem[You \etal\ (2007)]{yhc+07}
{You, X. P., et al.} 2007, \textsc{MNRAS}, 378, 493

\bibitem[Zhao (2011)]{zha11}
{Zhao, W.} 2011, \textsc{PhRvD}, 83, 4021

\end{thebibliography}
\end{document}